\documentclass[twocolumn]{aastex701}
\usepackage{amsmath}
\usepackage{booktabs}
\usepackage{multirow}
\usepackage{threeparttable}
\usepackage{makecell}
\usepackage{soul}
\usepackage{graphicx}
\usepackage{subcaption}
\usepackage{float}
\usepackage{xcolor}
\usepackage[normalem]{ulem}
\usepackage{cancel}
\usepackage{soul}

\begin{document}
\linenumbers

\title{Tracing Obscured AGN Contribution and Number Fraction
Across 0 $<$ z $<$ 6  with JWST}

\correspondingauthor{Angel Rodriguez Barbosa}
\email{s114025422@m114.nthu.edu.tw}

\author[0009-0009-6588-2271]{Angel Rodriguez Barbosa}
\affiliation{Institute of Astronomy, National Tsing Hua University, 101, Section 2, Kuang-Fu Road, Hsinchu, 30013, Taiwan}
\email{angelskyship@gmail.com}

\author[0000-0002-6821-8669]{Tomotsugu Goto}
\affiliation{Institute of Astronomy, National Tsing Hua University, 101, Section 2, Kuang-Fu Road, Hsinchu, 30013, Taiwan}
\affiliation{Department of Physics, National Tsing Hua University, 101, Section 2. Kuang-Fu Road, Hsinchu, 30013, Taiwan}
\email[]{}

\author[0000-0002-5687-0609]{Daryl Joe D. Santos}
\affiliation{Department of Physics, De La Salle University, 2401 Taft Ave, Malate, Manila, 1004 Metro Manila, Philippines}
\email[]{}

\author[0000-0002-3738-1834]{Chih-Teng Ling}
\affiliation{Department of Astronomical Science SOKENDAI, 2-21-1 Osawa, Mitaka, Tokyo, 181-8588, Japan}
\email[]{}

\author[0000-0001-7228-1428]{Tetsuya Hashimoto}
\affiliation{Department of Physics, National Chung Hsing University, 145, Xingda Road, Taichung, 40227, Taiwan}
\email[]{}

\author[0000-0002-9119-2313]{Ece Kilerci}
\affiliation{Department of Astronomy and Space Sciences, Istanbul University, Beyazıt, 34452 Fatih/İstanbul, Türkiye}
\email[]{}

\author[0000-0002-0524-5328]{Priyanka Jalan}
\affiliation{Institute of Astronomy, National Tsing Hua University, 101, Section 2, Kuang-Fu Road, Hsinchu, 30013, Taiwan}
\email[]{}

\author[0009-0009-2940-0861]{Terry Long Phan}
\affiliation{Institute of Astronomy, National Tsing Hua University, 101, Section 2, Kuang-Fu Road, Hsinchu, 30013, Taiwan}
\email[]{}

\author[0000-0002-5870-089X]{Deriyan Senjaya}
\affiliation{Department of Physics, National Tsing Hua University, 101, Section 2. Kuang-Fu Road, Hsinchu, 30013, Taiwan}
\email[]{}

\author[0009-0001-9195-7494]{Vignesh V.V. Rao}
\affiliation{Department of Physics, National Chung Hsing University, 145, Xingda Road, Taichung, 40227, Taiwan}
\email[]{}

\author[0009-0003-0031-0527]{Bahareh S. Salmasi}
\affiliation{Institute of Astronomy, National Tsing Hua University, 101, Section 2, Kuang-Fu Road, Hsinchu, 30013, Taiwan}
\email[]{}

\begin{abstract}

Active galactic nuclei (AGN) are key drivers of galaxy evolution, yet many remain undetected in ultraviolet and optical surveys due to heavy dust obscuration. In these systems, absorbed emission is re-radiated at infrared (IR) wavelengths, making IR observations essential for identifying the full AGN population. Tracking the AGN IR contribution and number fraction provides insight into both the dominance and prevalence of AGN activity across cosmic time. Using the JWST Systematic Mid-infrared Instrument Legacy Extragalactic Survey (SMILES) and the JWST Advanced Deep Extragalactic Survey (JADES), we leverage continuous optical-to-mid-IR coverage (0.4–25 $\mu$m) in the GOODS-S field to identify obscured AGN via multi-wavelength SED fitting with CIGALE. Our sample includes 278 AGN across 0 $<$ z $<$ 6, representing a seven-fold increase in sample size relative to previous studies utilizing the Cosmic Evolution Early Release Science (CEERS) survey due to the larger 15-pointing SMILES MIRI footprint. We find that both the AGN IR contribution and number fraction increase with redshift, with AGN fractions rising from $\lesssim$ 5\% at z $<$ 2 to $\sim$ 30\% at higher redshifts, while the median AGN contribution increases by up to $\sim$ 0.15. In contrast, as a function of total IR luminosity over $\log(L/L_{\odot}) \approx 8$--12, the AGN contribution and AGN number fraction remain fundamentally static. These trends suggest that the prevalence of obscured AGN activity is dependent on redshift while showing little to no dependence on total infrared luminosity.  Our results highlight JWST’s ability to uncover previously hidden AGN populations and provide new constraints on AGN--galaxy co-evolution.

\end{abstract}
\keywords{Active galactic nuclei (16), Infrared galaxies (790), Galaxy evolution (594), James Webb Space Telescope (2291)}

\section{Introduction} 
\label{sec:Intro}

Active galactic nuclei (AGNs) are compact, extremely luminous regions at the centers of galaxies, powered by the accretion of matter onto supermassive black holes (SMBHs). SMBHs are believed to reside in most massive galaxies and are thought to be closely linked to the evolution of their host systems across cosmic time \citep{ALEXANDER201293, annurev:/content/journals/10.1146/annurev-astro-082708-101811}. During their active phases, AGNs release large amounts of energy through radiation and outflows, which can influence star formation and regulate galaxy growth through feedback processes \citep{annurev:/content/journals/10.1146/annurev-astro-081811-125521, annurev:/content/journals/10.1146/annurev-astro-081913-035722}. Because of this, understanding the demographics and evolution of AGNs is an important step toward building a complete picture of galaxy evolution.

AGNs can be identified through a variety of observational techniques across the electromagnetic spectrum. Optical spectroscopy is commonly used to distinguish AGNs from star-forming galaxies using emission-line diagnostics such as the BPT diagram \citep{Baldwin_1981, Kewley_2001, 10.1111/j.1365-2966.2003.07154.x}, while X-ray and radio observations provide additional methods for identifying AGN activity that is less affected by dust obscuration \citep{1971ApJ...165L..43G, 2016A&ARv..24...13P, Hickox2018ARA&A..56..625H}. However, each of these methods is subject to selection biases, and a significant population of AGNs, particularly those that are heavily dust obscured, can remain undetected in traditional surveys; meaning the true fraction of these obscured AGN is still debated \citep{2013ApJ...770...40M, 10.1093/mnras/stv2748}. Multi-wavelength censuses in deep fields like GOODS-S have highlighted these selection gaps, demonstrating that individual selection regimes (whether optical, X-ray, or radio) frequently miss distinct sub-populations of active systems \citep{2022ApJ...941..191L}. While wide-area X-ray surveys confirm that the obscured AGN fraction rises significantly up to $z \sim 3$ \citep{Peca_2023}, heavily buried, Compton-thick systems remain difficult to identify.

In these obscured systems, ultraviolet (UV) and optical emission from the accretion disk is absorbed by surrounding dust, often associated with the torus, and re-emitted at infrared (IR) wavelengths. As a result, infrared observations provide a powerful tool to probe AGN activity that is relatively insensitive to dust extinction. In particular, mid-infrared (mid-IR) wavelengths are especially useful for separating AGN emission from that of star formation, as they trace hot dust heated by the central engine \citep{2017ApJ...849..111K, Kirkpatrick2023ApJ...959L...7K}. This mid-IR emission is physically driven by dust grains surrounding the central engine that absorb the primary optical and UV accretion energy and thermally re-radiate it as a hot continuum peaking at mid-infrared wavelengths \citep{1987ApJ...320..537B}. Consequently, multi-wavelength surveys rely heavily on mid-IR selection as a highly complete method to identify obscured active nuclei that are frequently missed by optical or X-ray criteria due to severe obscuration \citep{Hickox2018ARA&A..56..625H}.

Despite this, studies prior to the launch of the James Webb Space Telescope (JWST) were limited by incomplete wavelength coverage beyond $\sim$8 $\mu$m, which introduced degeneracies in spectral energy distribution (SED) fitting and made it difficult to robustly disentangle AGN and host galaxy contributions. Early space-based infrared observations, particularly with the Spitzer Space Telescope, successfully demonstrated that obscured active galaxies could be efficiently identified using distinct mid-infrared color signatures \citep{2004ApJS..154..166L}. However, these earlier missions lacked the continuous wavelength coverage and sensitivity required to detect faint, high-redshift sources.

With the advent of JWST, we are now seeing a major shift in how AGNs are studied. Using NIRSpec and NIRCam observations at $\lambda$ $<$ 5 $\mu$m, recent studies have identified a growing number of high-redshift AGNs and AGN candidates through rest-frame UV–optical emission line diagnostics, narrow highionization lines, and compact red colors \citep{2023A&A...677A.145U, 2023ApJ...959...39H, 2023ApJ...954L...4K, 2023ApJ...953L..29L, 2025A&A...697A.175S, 2026MNRAS.546ag086J}, with detections extending to z $\sim$ 10 \citep{2023ApJ...955L..24G, 2024Natur.627...59M}. However, because these near-infrared selections fundamentally rely on rest-frame ultraviolet and optical signatures at higher redshifts, they remain inherently limited by dust attenuation which can severely suppress or completely extinguish emission line features.\citep{2026MNRAS.546ag086J, 2025A&A...697A.175S} Moreover, some of these searches tend to select broad-line, unobscured systems, and therefore may not represent the full AGN population. As a result, the true number of AGNs, particularly those that are heavily obscured, remains uncertain.

Previous systematic searches for obscured AGNs in the infrared have also been limited by the small number of available photometric bands beyond $\sim$8 $\mu$m, leaving SED models unchecked. This limitation has now been largely overcome with JWST’s Mid-Infrared Instrument (MIRI), which provides eight continuous photometric bands spanning 5.65–25.5$\mu$m observing the hot dust emission associated with AGN activity ($\lambda_{\text{peak}}$ $\ge$3 $\mu$m) out to z $\sim$ 8 \citep{2024ApJ...966..229L}. Compared to earlier missions such as Spitzer Space Telescope or AKARI, MIRI achieves $\sim$90× higher sensitivity and significantly improved spatial resolution, enabling the detection and characterization of AGNs across a wider range of redshift, luminosity, and obscuration \citep{Yang2023ApJ...950L...5Y}

Recent studies using JWST surveys have already identified a substantial population of new dust-obscured AGNs, highlighting JWST as a key facility for investigating obscured AGN. For instance, \cite{2024ApJ...966..229L} utilized mid-IR SED modeling with the SMILES survey to find that they could recover heavily dust obscured, Compton-thick AGN. Approximately 80\% of MIRI-selected AGN candidates represented new discoveries that had been missed by previous deep X-ray or optical searches. Furthermore, recent broad mid-IR background studies show a steady increase in the overall obscured AGN fraction
extending all the way from z = 0 up to z = 6 This shift marks a new era in completing the cosmic census of black hole growth, particularly for the most heavily obscured populations \citep{bulichi2026meowincreaseobscuredagn}.


Previous studies such as \cite{Wang2020MNRAS.499.4068W} and \cite{Chien2024MNRAS.532..719C}, reported that AGN contribution and number fraction increase with redshift, while occasionally decreasing with higher total IR luminosity. However, many of these works were limited by incomplete mid-IR coverage, which can introduce degeneracies between AGN and star-formation radiation, or lacked sufficient high redshift data. The present study addresses this limitation by utilizing the JWST SMILES survey; its continuous MIRI coverage facilitates more accurate SED fitting and a more robust separation between AGN and star-forming galaxies (SFGs). Consequently, this dataset provides a more reliable basis for testing previously reported trends in AGN demographics.

In this work,we investigate how AGN IR contribution (labeled $frac_{AGN}$), defined as the ratio between AGN infrared luminosity and total infrared luminosity of the host galaxy, and AGN number fraction evolve with redshift and total IR luminosity using JWST's SMILES in conjunction  with the JADES datasets in the GOODS-S field. By utilizing the full optical to mid-IR SED fitting results from \cite{2026PASP..138a4102L}, we examine these relations with a much larger and more sensitive sample. Our goal is to investigate if the previously reported trends remain valid when using JWST’s continuous mid-IR coverage, and how JWST-selected AGNs change or expand our understanding of obscured AGN demographics.

This paper is structured as follows. In Section \ref{sec:Data}, we describe the datasets used in this study, including the JWST SMILES and JADES surveys, as well as the SED fitting procedure and sample selection. In Section \ref{sec:Results}, we present our main results, focusing on the evolution of AGN IR contribution and AGN number fraction as functions of redshift and total IR luminosity. In Section \ref{sec:Discussion}, we discuss the implications of these results and compare them with previous studies. Finally, in Section \ref{sec:Conclusion}, we summarize our main conclusions and highlight potential directions for future work.

Throughout the paper, we adopt the Planck18 cosmology \citep{Planck2020A&A...641A...6P} with ($\Omega_{m}$, $\Omega_{\Lambda}$, $\Omega_{b}$, $h$) = (0.310, 0.689, 0.0490, 0.677)

\begin{figure*}[ht]
    \centering
    \begin{subfigure}{0.45\textwidth}
        \includegraphics[width=\textwidth]{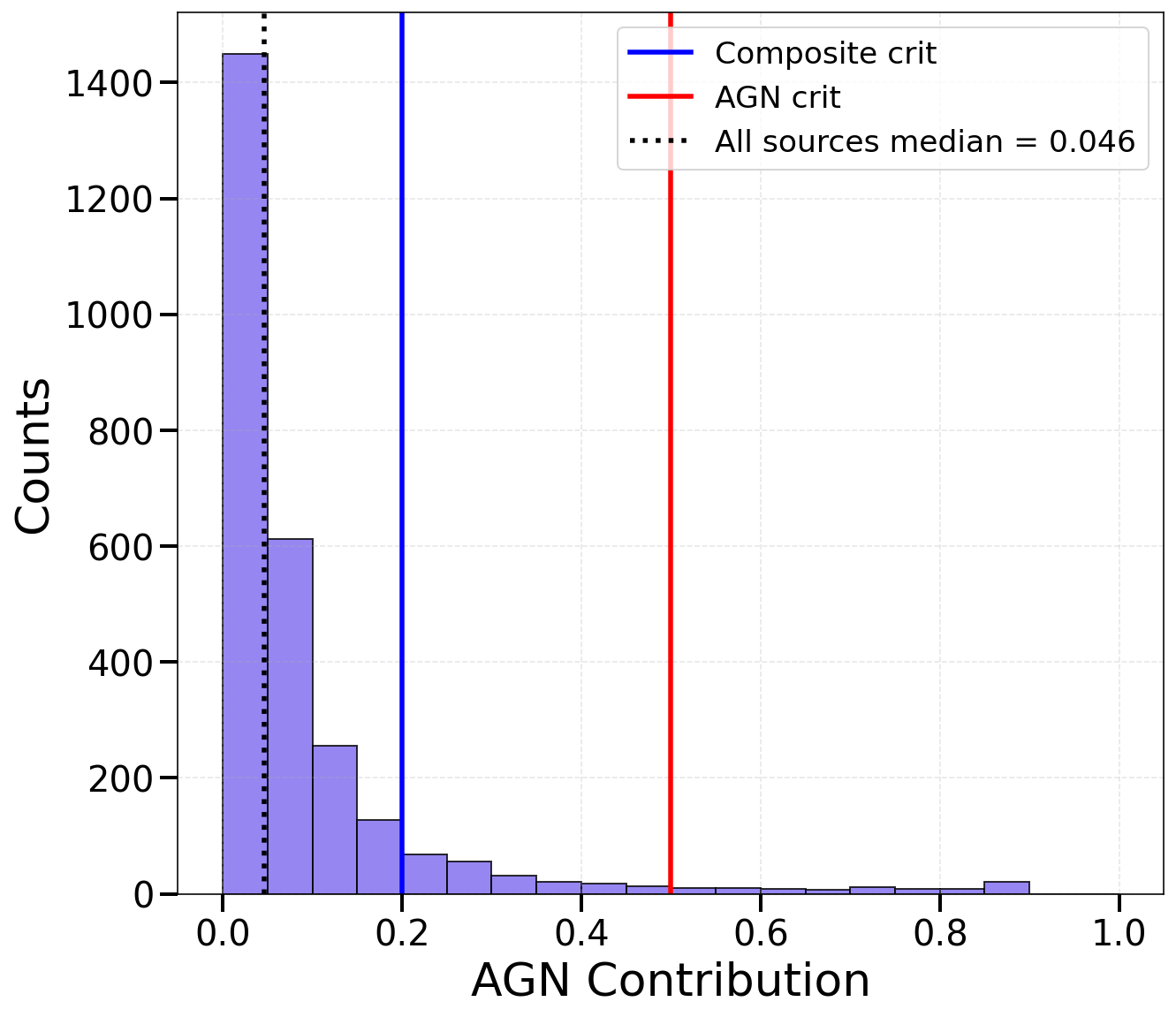}
        \caption{}
    \end{subfigure}
    \hfill
    \begin{subfigure}{0.45\textwidth}
        \includegraphics[width=\textwidth]{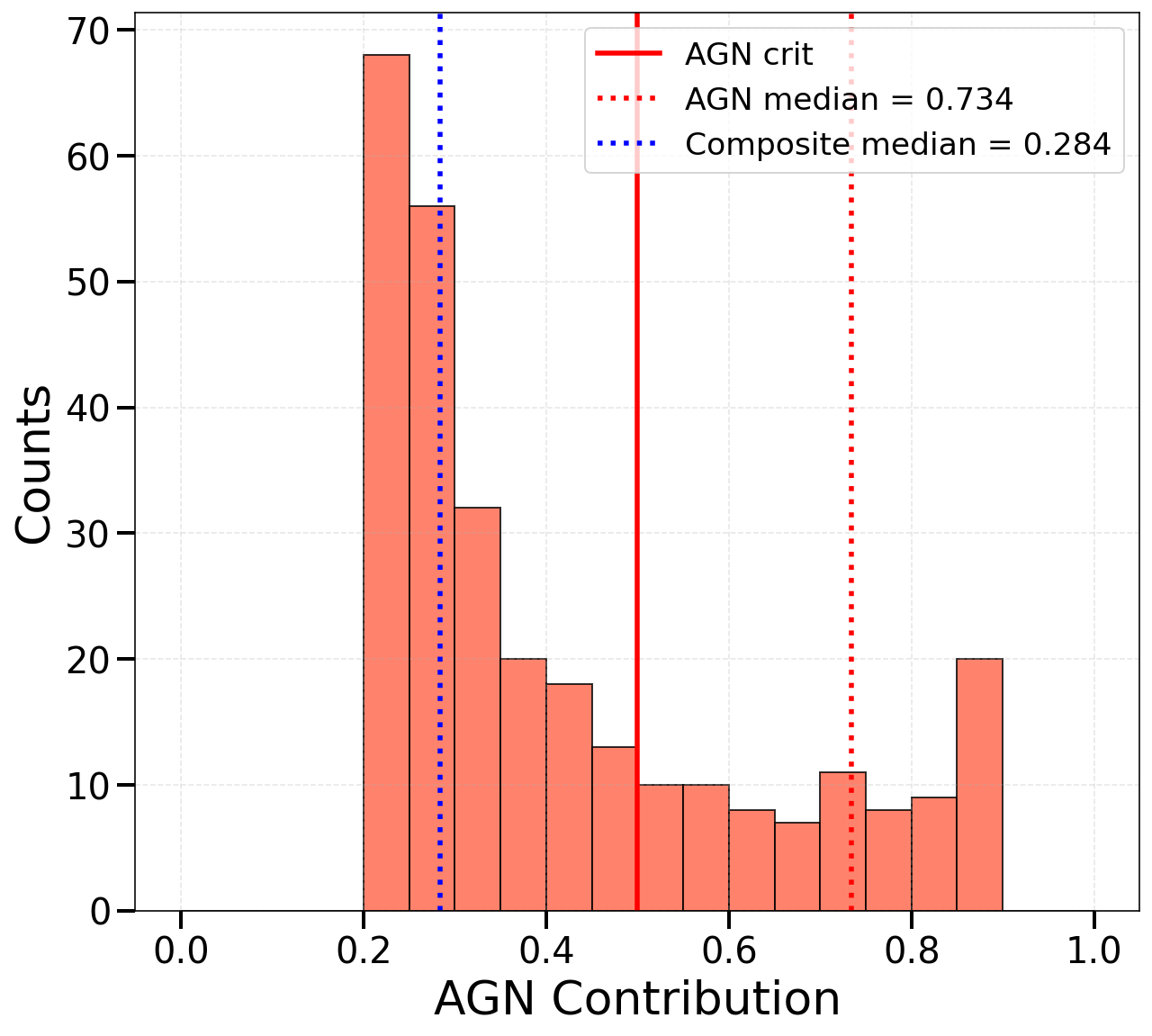}
        \caption{}
    \end{subfigure}
    \caption{(a) Histogram showing the distribution of all sources by AGN Contribution. The solid lines marking the criteria for composite (0.2 $\le frac_{AGN} <$ 0.5) and AGN ($frac_{AGN} \ge$ 0.5) sources, while the dotted line shows the total median value. (b) Zoomed in area showing clearer distribution of our composite and AGN sources. Solid line again showing the criteria for AGN, while the dotted lines show the median value for AGN and composite sources}
    \label{fig:Hist1}
\end{figure*}

\begin{figure*}[ht]
    \centering
    \includegraphics[width=0.7\textwidth]{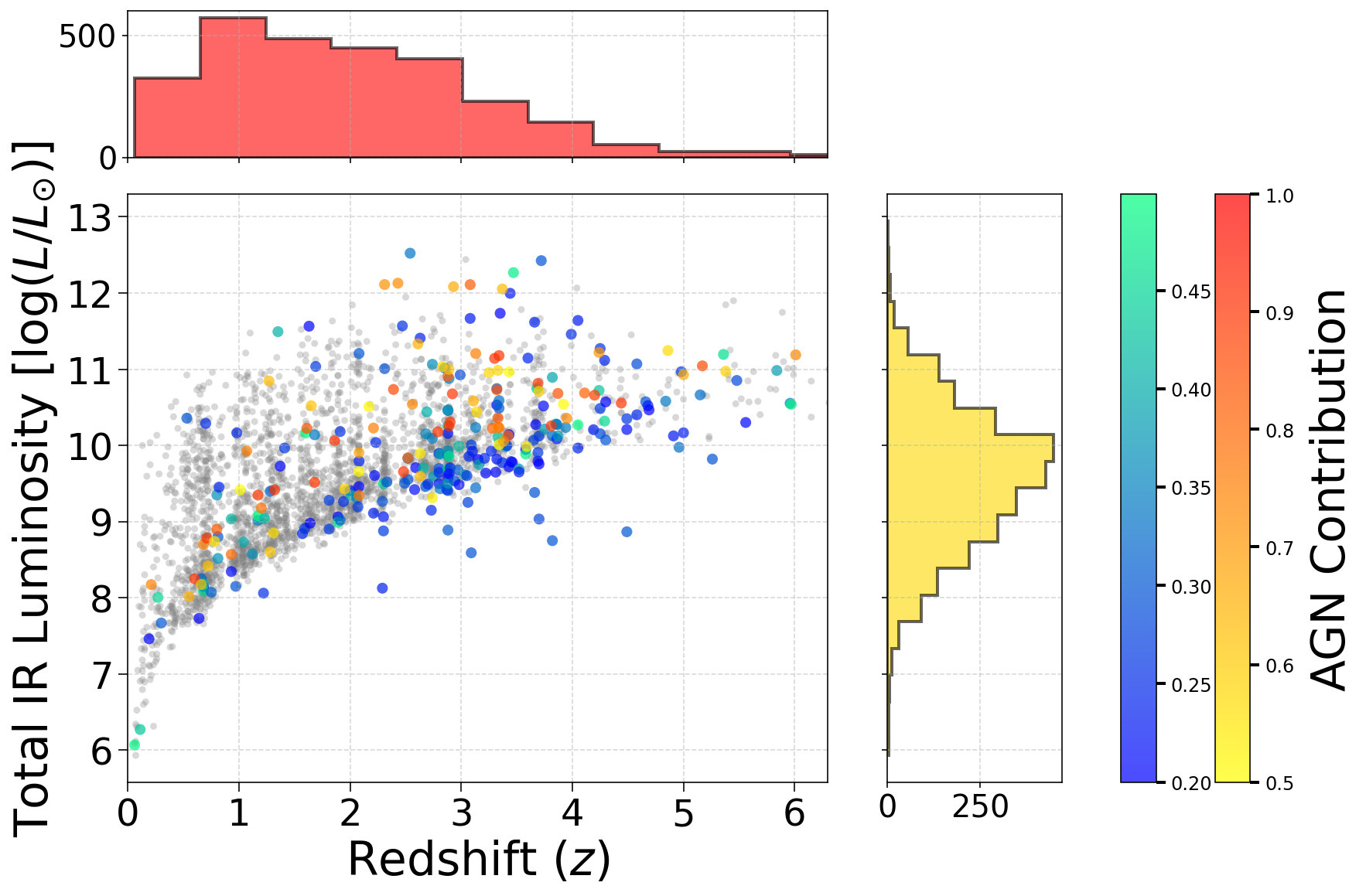}
    \caption{Distribution of all galaxies in total infrared luminosity and redshift. The main panel shows $\log(L_{\mathrm{TIR}}/L_\odot)$ as a function of redshift ($z$). Galaxies with $frac_{\mathrm{AGN}}$ $<$ 0.2 are shown in gray, while composite and AGN sources are color-coded according to their AGN fractional contribution ($frac_{\mathrm{AGN}}$ $\geq$ 0.2). The top and right panels show the redshift and luminosity distributions of the sample, respectively.}
    \label{fig:agn_ltir_redshift}
\end{figure*}

\section{Data} \label{sec:Data}

\subsection{SMILES} 
The JWST Systematic Mid-infrared Instrument Legacy Extragalactic Survey \citep[SMILES, PID 1207;][]{SMILES_Rieke2024} provides continuous observation in the mid-IR with JWST's 8 MIRI bands ranging from 5.6 - 25.5$\mu$m (F560W, F770W, F1000W, F1280W, F1500W, F1800W, F2100W, F2550W). The survey covers an area of $\sim$ 34 arcmin$^2$ in the GOODS-S/HUDF field, with an exposure time of approximately 650 - 2100 seconds per band. SMILES currently provides the most complete continuous MIRI coverage available over a deep extragalactic field; minimizing degeneracies in spectral energy distribution (SED) fitting, and providing a more rigorous framework for studying mid-IR galaxy populations.

This study uses the photometric catalog from the SMILES initial data release \citep{SMILES_Alberts2024}. By aligning the SMILES imaging to the JADES Nircam catalog \citep{SMILES_Rieke2024}, the astrometric accuracy is found to be 0.1 - 0.2 (0.4) arcsec for F560W-F2100W (F2550W). To perform SMILES photometry, a modified version of the JADES photometric pipeline specifically designed for MIRI data is used \citep{JADES_Rieke2023ApJS..269...16R}. \cite{SMILES_Alberts2024} use the combined F560W+F770W stack detection image to identify sources and measure their aperture photometry in all eight MIRI bands using circular and Kron apertures. The resulting catalog contains 3096 sources with $\ge$ 4$\sigma$ detection. 

While traditional mid-IR AGN identification often relies on discrete color-selection such as the diagnostics developed in \cite{2017ApJ...849..111K}; these methods are susceptible to degeneracies between AGN emission and host galaxy star formation. These degeneracies arise primarily from host galaxy contamination of the 1.6 $\mu$m stellar bump, redshift-induced filter shifts, and the overlap of PAH and silicate features with broadband filters. The SMILES survey mitigates these issues by providing continuous coverage across eight MIRI filters (5.6–25.5 $\mu$m). This high density sampling enables full SED decomposition, breaking the ambiguity of composite systems and providing a more robust separation of AGN-heated dust from star-forming features than static color thresholds.

\subsection{Additional Data}
While mid-IR photometry is critical for identifying obscured accretion, it remains insufficient in isolation to study the physical properties of AGNs; datasets of multiple more wavelengths are needed for accurate SED fitting and to effectively distinguish sources between AGNs and SFGs. The catalog from the JWST Advanced DEEP Extragalactic Survey \citep[JADES, PID 1180;][]{JADES_2023arXiv230602465E} Data Release 3 \citep{JADES_Rieke2023ApJS..269...16R, JADES_2024A&A...690A.288B, JADES_DR3_2024arXiv240406531D} located in the GOODS-S field complements SMILES perfectly, providing deep near-IR imaging in 9 NIRcam bands (F090W-F444W), including additional medium band filters (F182M, F210M, F4309M, F460M, F480M) from the JWST Extragalactic Medium band Survey \citep[JEMS, PID 1963][]{JEMS_2023ApJS..268...64W}. JADES also includes 9 other bands combining HST / WFC3 IR and ACS based on data from the CANDLES survey, resulting in 23 optical and near-IR bands in total. Together, SMILES and JADES provide 23 optical–near-IR bands and 8 mid-IR bands, enabling robust optical-to-mid-IR (0.4-25.5 $\mu$m) SED fitting .

\subsection{SED Fitting and Sample Selection}
We use the SED fitting results presented in \cite{2026PASP..138a4102L}, who constructed the crossmatched SMILES+JADES catalog mentioned above using 8 MIRI bands together with 23-band HST+JWST/NIRCam photometry, derived photometric redshifts with EAZY, which are then used to perform optical-to-mid-IR SED fitting with CIGALE v2022.1.

EAZY takes user-supplied galaxy SED templates to fit observed photometry and determine the best-fit redshift. In their study, they allow EAZY photometric redshifts a range of 0-10, and found that EAZY photometric redshifts with NIR-only data from JADES was more reliable than NIR + mid-IR data from JADES and SMILES; perhaps due to the relatively high uncertainties of MIRI at longer wavelengths and incomplete mid-IR templates in EAZY.

After collecting the redshifts using EAZY, they performed optical-to-mid-IR SED fitting using CIGALE. Briefly put, their CIGALE configuration included the modules:
\begin{itemize}
    \item \textbf{sfhdelayed:} Models a galaxy's star formation rate, ($\text{SFR} \propto t \exp(-t/\tau)$) characterized by a stellar population age ($t$) and an e-folding decay time ($\tau$), as a gradual delayed rise to a peak followed by an exponential decline to handle long-term galaxy evolution; with values $\tau = [0.5, 1, 2, 4]$~Gyr.
 
    \item \textbf{bc03:} Utilizes \cite{2003MNRAS.344.1000B} stellar population synthesis templates  to generate the base unattenuated stellar spectrum.
    
    \item \textbf{nebular:} Simulates the emission from gas ionized by massive stars. It adds both nebular emission lines and the nebular continuum to the stellar spectrum.
    
    \item \textbf{dustatt\_modified\_CF00:} Implements a modified \cite{2000ApJ...539..718C} attenuation model that calculates how much starlight is absorbed and scattered by interstellar dust, and subsequently determines how much energy is re-emitted by that dust; with the visual attenuation grid spanning $A_V^{\text{ISM}} = [0.01, 0.02, 0.04, 0.08, 0.16, 0.32, 0.63, 1.3, 2.5, 5, 10]$ mag.
    
    \item \textbf{dl2014:} Models the galactic dust grain emission heated by star formation based on the \citep{2014ApJ...780..172D} templates, tracking polycyclic aromatic hydrocarbon (PAH) mass fractions and dust grain heating.
    
    \item \textbf{skirtor2016:} Simulates the active galactic nucleus (AGN) component via a 3D clumpy torus template calculated from the \texttt{SKIRT} radiative transfer code \citep{2016MNRAS.458.2288S}. The torus configuration assumes a fixed edge-on viewing angle of $70^\circ$ and an average optical depth at $9.7\,\mu\text{m}$ varying between $\tau_{9.7} = [3, 5, 7, 9, 11]$, and an AGN infrared contribution step grid spanning $f_{\text{AGN}} = [0, 0.01, 0.03, 0.05, 0.1, 0.2, 0.3, 0.5, 0.75, 0.9]$.
\end{itemize}

Our final cross-matched SED catalog contains 2,735 sources. From this parent sample, we extract the derived photometric redshifts, AGN IR luminosities ($L_{AGN}$), and AGN IR contributions ($frac_{AGN}$) to carry out our analysis. By scaling the skirtor2016 AGN torus templates against the host galaxy dust and stellar components, CIGALE successfully separates the starburst emission from the central black hole activity.

Within this framework, the AGN IR contribution ($frac_{AGN}$) is calculated as the ratio between the AGN IR luminosity ($L_{AGN}$) and the total IR luminosity of the galaxy ($L_{TIR}$), integrated across a rest-frame mid-IR window of $3\text{--}30\,\mu\text{m}$. This parameter describes how much the active nucleus contributes to the total IR emission of the host galaxy, serving as our primary metric for separating star-forming dominated galaxies and AGN dominated galaxies.

We then use these calculated $frac_{AGN}$ values to determine each galaxy’s type. Following the classification thresholds and conventions established in \cite{Wang2020MNRAS.499.4068W, Chien2024MNRAS.532..719C}, which we refer to as Equation \eqref{eq:AGN_crit}

\begin{align}
\label{eq:AGN_crit}
    {\ frac_{AGN}} = \frac{L_{AGN}}{L_{TIR}} \ge 0.2
\end{align}

as our criteria for classifying AGNs. In other words, the AGN luminosity should contribute at least 20\% of the total galaxy luminosity for a source to be labeled as having AGN activity. We follow the same criterion set by \cite{Wang2020MNRAS.499.4068W}, dividing our candidates into composite sources (0.2 $\le frac_{AGN} <$ 0.5), AGN ($frac_{AGN} \ge$ 0.5), and SFG ($frac_{AGN} < 0.2$). From this we find 83 AGNs, 207 composites, and 2445 SFGs out of the 2735 total sources analyzed by CIGALE. Figure \ref{fig:Hist1} and Figure \ref{fig:agn_ltir_redshift} display the distribution of these sources.

We note that due to the lack of far-infrared (FIR) observations at comparable spatial resolution (e.g., from Herschel or AKARI), our SED fitting around the FIR peak where dust emission dominates relies primarily on constraints from shorter mid-IR wavelengths. This limitation can introduce uncertainties in the derived total IR luminosity , especially at higher redshifts. A more detailed discussion of these effects is presented in \cite{Ling2024}. However, for SFGs, previous studies have shown that there exists a strong empirical relationship between mid-IR luminosity and total IR luminosity, allowing mid-IR observations to serve as a reliable proxy for the full IR SED \citep{Caputi2007,Goto2011MNRAS.410..573G}. Therefore, despite the lack of direct FIR constraints, our SED fitting remains robust for the purposes of this study.

\section{Results} \label{sec:Results}

\subsection{AGN Contribution}\label{sec:style}
We investigate whether there is a link between AGN contribution and redshift or total IR luminosity. For each redshift and luminosity bin, we calculate the median AGN contribution using the full galaxy sample, including SFGs, composite sources, and AGNs. The resulting trends reflect the overall importance of obscured AGN activity within the galaxy population as a whole with respect to redshift and IR luminosity.

\begin{figure*}[ht]
    \centering
    \begin{subfigure}{.49\textwidth}
        \includegraphics[width=\textwidth]{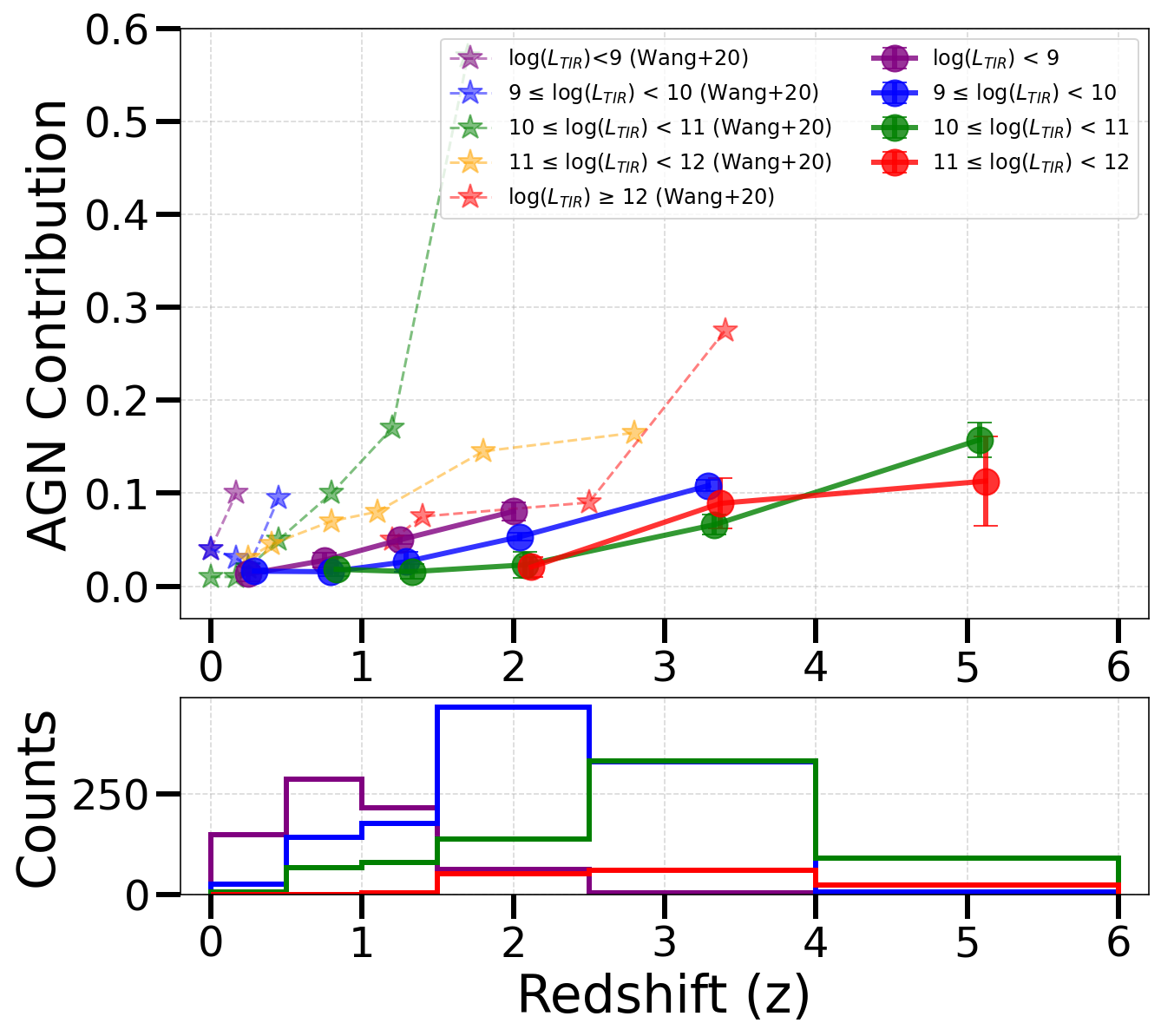}
        \caption{}
        \label{subfig:AGN_contA}
    \end{subfigure}
    \hfill
    \begin{subfigure}{0.49\textwidth}
        \includegraphics[width=\textwidth]{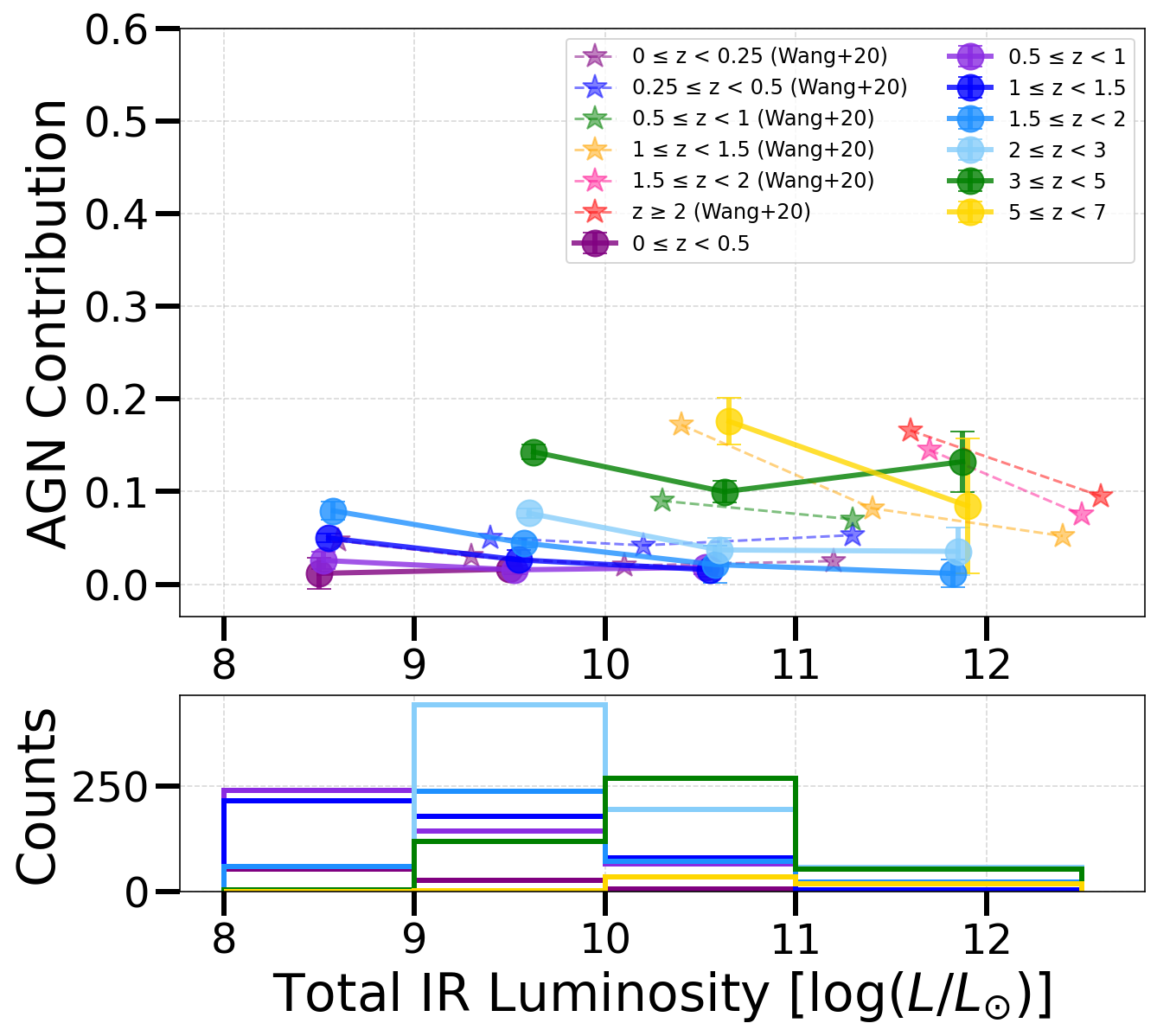}
        \caption{}
        \label{subfig:AGN_contB}
    \end{subfigure}
    \caption{(a) AGN contribution as a function of redshift. A different color is used for each increasing luminosity subsample. We also add a small horizontal offset to each line for easier visibility. Bottom panel shows the redshift distribution of sources within each luminosity subsample. Note that in this figure we perform dynamic increasing binning; meaning that at lower redshifts we have smaller bin widths, and at higher redshifts we have longer bin widths. The reason for this is because we need to take into consideration the timescales across each redshift to have a robust statistical sample in each interval. (b) AGN contribution as a function of total IR luminosity. A different color is used for each increasing redshift subsample;  and we add a small horizontal offset to each line for easier visibility. Bottom panel shows the luminosity distribution of sources within each redshift subsample.}

    \label{fig:AGN_Cont}
\end{figure*}

Figure \ref{subfig:AGN_contA} shows AGN contribution as a function of redshift. To account for the timescales across each redshift and ensure a robust statistical sample in each interval, we organize our bins from small to larger redshift bins. We divide our redshifts into 6 bins bounded by z = [0.0, 0.5, 1.0, 1.5, 2.5, 4.0, 6.0], and add a small horizontal offset for easier visibility. For each bin, we report the median AGN contribution to provide a robust measure of the typical AGN contribution that is less sensitive to outliers. The associated uncertainty was estimated from the scatter of the individual AGN contribution values within each bin as $\frac{\sigma}{\sqrt{N}}$, where $\sigma$ is the standard deviation of the AGN contribution values and N is the number of galaxies in the bin. This approach quantifies the uncertainty in the AGN contribution arising from the intrinsic scatter of the galaxy population within each bin. The error bars therefore reflect the statistical uncertainty in the median AGN contribution of each galaxy population, rather than the measurement uncertainty of individual SED fits. To ensure statistically meaningful measurements, bins containing fewer than seven galaxies were excluded from the analysis. This threshold was adopted to minimize the influence of low-number statistics and to avoid over-interpreting measurements derived from sparsely populated bins. The same criterion was applied consistently to both the AGN contribution and AGN number fraction analyses.

We find a positive trend of AGN contribution across all luminosity subsamples with increasing redshift, which is consistent with \cite{Wang2020MNRAS.499.4068W} and \cite{Chien2024MNRAS.532..719C}. We find our lowest luminosity subsample, log($L_{TIR}$) $<$ 9, show the with the smallest increase in AGN contribution from 0.01 at z = 0.25 to 0.08 at z = 2. Our intermediate subsamples of 9 $\leq$ log($L_{TIR}$) $<$ 10 and 10 $\leq$ log($L_{TIR}$) $<$ 11, show a more dramatic increase from $\sim$ 0 at z $<$ 1, to 0.12 at z = 3.25 and 0.16 at z = 5 respectively. While our most luminous subsample, 11 $\leq$ log($L_{TIR}$) $<$ 12 increases in AGN contribution starting from 0.02 at z = 2 to 0.11 at z = 5.

Figure \ref{subfig:AGN_contB} illustrates AGN contribution as a function of total IR luminosity. To evaluate our trends, we divide our luminosities into 4 bins bounded by log($L/L_{\odot}$) = [8, 9 , 10, 11, 12.5]. The size of the final bin is expanded to a width of 1.5 to account for the low number density of high-luminosity sources in our sample, ensuring a statistically robust analysis.

We do not observe a clear increase or decrease in AGN contribution with increasing total IR luminosity. Across most redshift subsamples, the trends remain broadly consistent with being flat within the uncertainties, although some groups exhibit mild variations. Our lower redshift subsamples, spanning 0 $\leq$ z $<$ 2, remain approximately flat across the entire luminosity range. The 3 $\leq$ z $<$ 5 subsample presents a complex trend, showing a slight decrease in AGN contribution starting from 0.14 at log($L/L_{\odot}$) = 9.5 to 0.1 at log($L/L_{\odot}$) = 10.5 and finally a slight increase to 0.13 at log($L/L_{\odot}$) = 11.75. While our highest redshift subsample, 5 $\leq$ z $<$ 7, shows a decrease in AGN contribution from 0.18 at log($L/L_{\odot}$) = 10.5 to 0.08 at log($L/L_{\odot}$) = 11.75.

These results may suggest that AGN generally contribute more significantly to the total IR photon budget at higher redshift, while having a weak or no dependence on luminosity.

\subsection{AGN Number Fraction}

\begin{figure*}[ht]
    \centering
    \begin{subfigure}{.49\textwidth}
        \includegraphics[width=\textwidth]{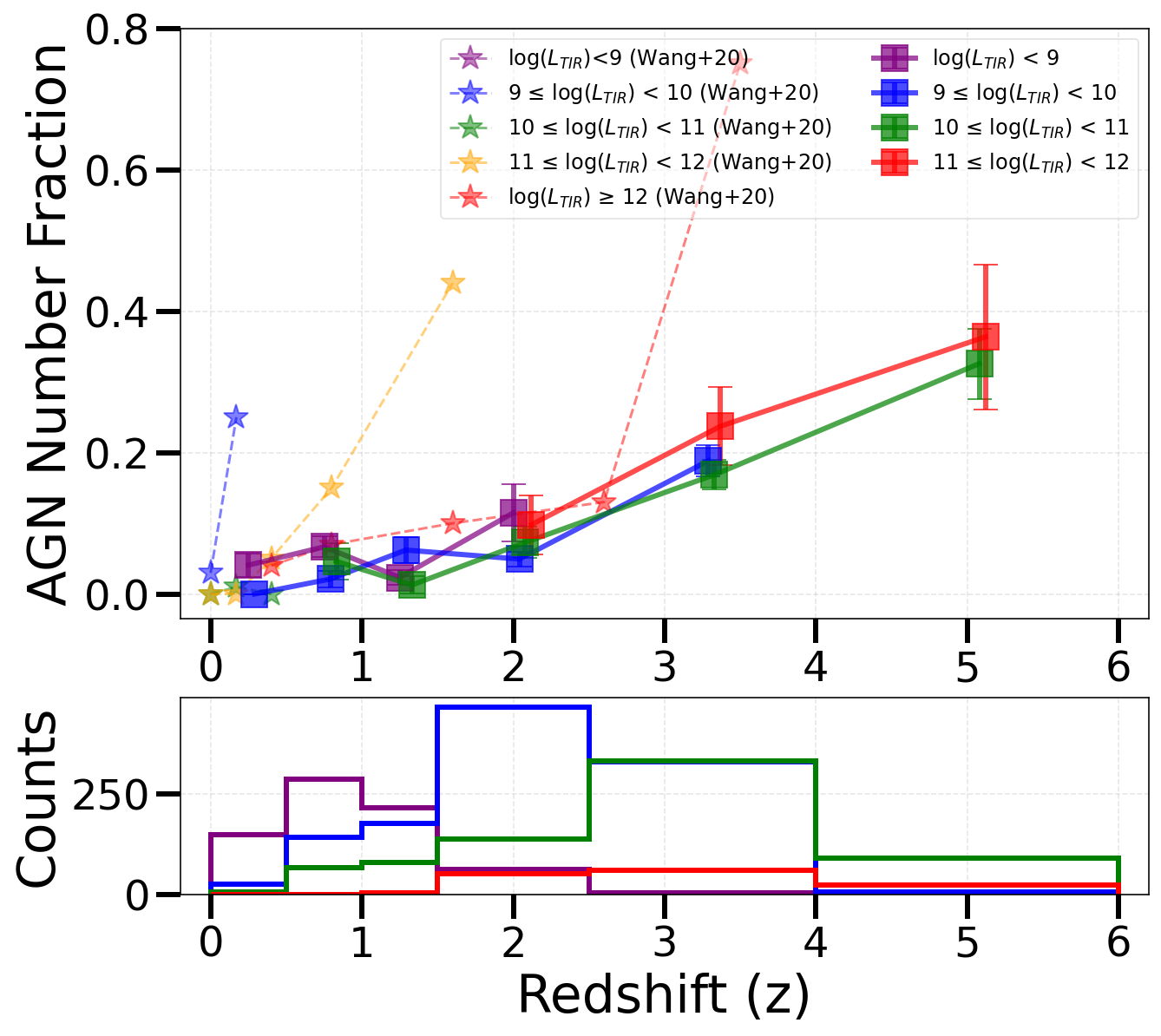}
        \caption{}
        \label{subfig:AGN_numfracA}
    \end{subfigure}
    \hfill
    \begin{subfigure}{0.49\textwidth}
        \includegraphics[width=\textwidth]{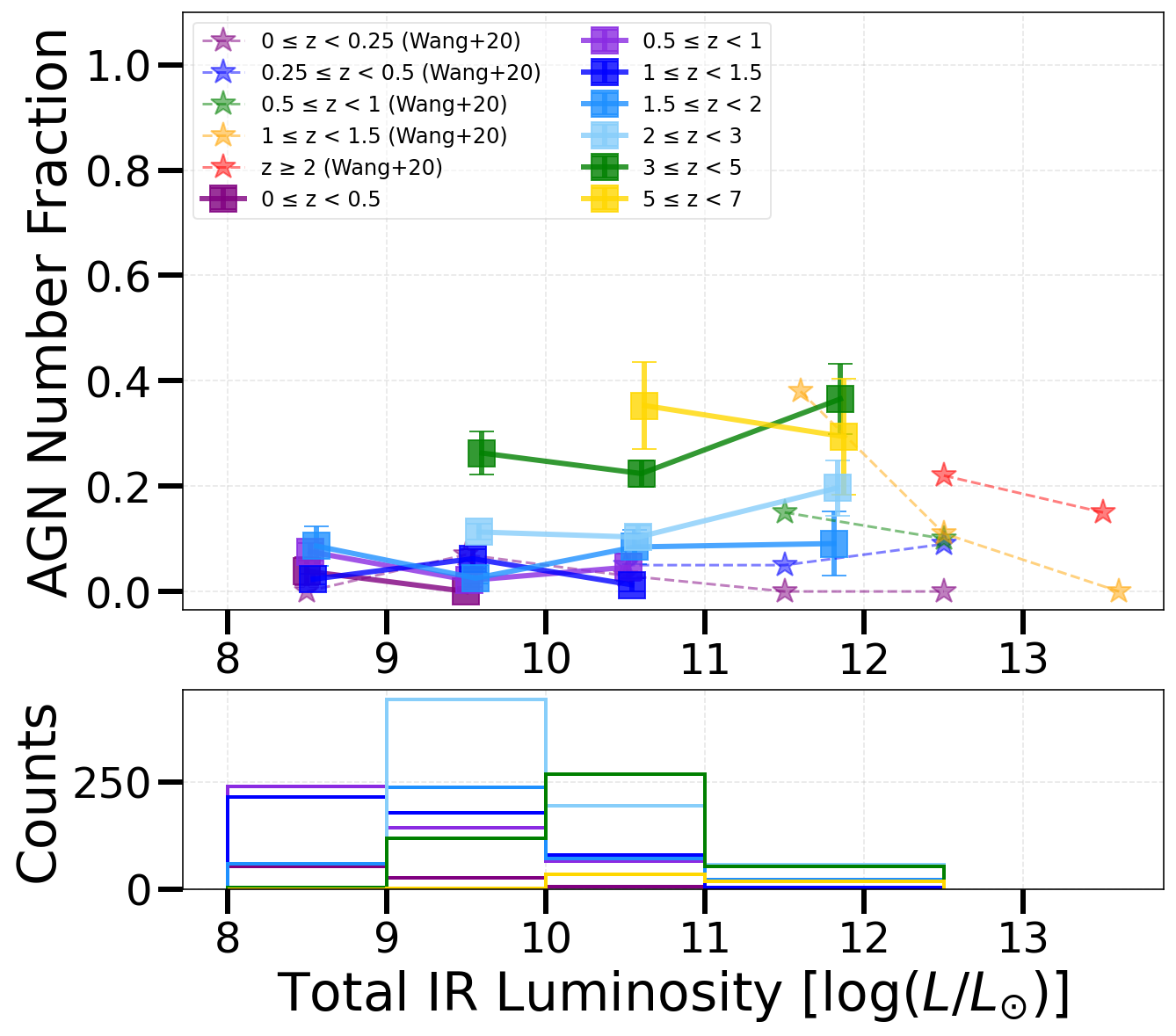}
        \caption{}
        \label{subfig:AGN_numfracB}
    \end{subfigure}
    \caption{(a) AGN number fraction as a function of redshift. A different color is used for each increasing luminosity subsample, and we add a small horizontal offset to each line for easier visibility. Note that we use increasing redshift bin widths to account for the timescales across each redshift for a robust statistical sample in each interval. Bottom panel shows the redshift distribution of sources within each luminosity subsample. (b) AGN number fraction as a function of total IR luminosity, and we add a small horizontal offset to each line for easier visibility. A different color is used for each increasing redshift subsample. Bottom panel shows the luminosity distribution of sources within each redshift subsample.} 
    \label{fig:AGN_NumFrac}
\end{figure*}

The other parameter that we are interested in is AGN number fraction ($f_{num}$), and how it changes with respect to redshift and total IR luminosity; it can be used to estimate how AGNs occupy certain redshift or luminosity bins and with what distribution. The definition of AGN number fraction is presented in Equation \eqref{eq:AGN_num_frac}:

\begin{align}
\label{eq:AGN_num_frac}
    {\ f_{num}} = \frac{N_{AGN} +N_{comp} }{N_{tot}} 
\end{align}

where $N_{AGN}$ is the total number of AGN sources, $N_{comp}$ refers to the total number of composite sources, and $N_{tot}$ refers to the sum of all total sources; including AGNs, composites, and SFGs. 

Figures \ref{subfig:AGN_numfracA} and \ref{subfig:AGN_numfracB} show our results. We calculate AGN number fraction in each bin, and the uncertainty on the AGN number fraction in each bin was estimated assuming binomial counting statistics, $\sigma = \sqrt{\frac{f(1-f)}{N}}$ , where $f$ is the measured AGN number fraction and $N$ is the total number of galaxies within the bin. This accounts for the statistical uncertainty that arises from measuring a fraction using a finite number of galaxies.

Figure \ref{subfig:AGN_numfracA} presents AGN number fraction as a function of redshift where we use the same horizontal offset and redshift bin widths as Figure \ref{subfig:AGN_contA}, and we observe a positive trend. Our faintest subsample, log($L_{TIR}$) $\le$ 9, sees the smallest change going from 0.04\% at z = 0.25 and increasing to 0.11\% at z = 2. However, our other luminosity subsamples see a more dramatic change over redshift starting at near zero number fraction below z = 2, and increasing to as much as 18\% at z = 3.25 for our 9 $\leq$ log($L_{TIR}$) $<$ 10 subsample, 33\% at z = 5 for our 10 $\leq$ log($L_{TIR}$) $<$ 11 subsample, and  36\% at z = 5 for our most luminous subsample, 11 $\leq$ log($L_{TIR}$) $<$ 12. These trends are consistent with \cite{Chien2024MNRAS.532..719C, Wang2020MNRAS.499.4068W}.  

Figure \ref{subfig:AGN_numfracB} illustrates AGN number fraction as a function of total IR luminosity, where we use the same bin widths as Figure \ref{subfig:AGN_contB}. We do not find strong statistical evidence for a significant decrease or increase in AGN number fraction with increasing total IR luminosity, as the trends are generally consistent with being flat within the uncertainties, although some redshift subsamples may show weak variations. The lower redshift subsamples spanning 0 $\leq$ z $<$ remain approximately constant, across the whole luminosity range. Our highest redshift subsample, 5 $\leq$ z $<$ 7, shows a slight decrease from 35\%  at log($L/L_{\odot}$) = 10.5 to 30\% at log($L/L_{\odot}$) = 11.5. The intermediate redshift subsample however, 3 $\leq$ z $<$ 5, portrays a more complex trend where we see a slight decrease from 26\% to 22\%, and then a slightly bigger increase to 32\% all taking place in the luminosity range of 9.5 $<$ log($L/L_{\odot}$) $<$ 11.5.

\section{Discussion} \label{sec:Discussion}

In this study, our objective was to investigate how AGN IR contribution and AGN number fraction evolve with redshift and total IR luminosity using the SMILES and JADES datasets from JWST. Our results show that both AGN contribution and AGN number fraction increase with redshift, suggesting that obscured AGN activity becomes more prevalent and contributes more significantly to the total IR emission in higher-redshift environments. In contrast, neither AGN contribution nor AGN number fraction shows strong statistical evidence for a dependence on total IR luminosity, as both trends remain broadly consistent with being flat within the uncertainties. These results suggest that redshift may play a more significant role than total IR luminosity in governing the prevalence and relative importance of obscured AGN within our sample.

Increasing AGN contribution as a function of redshift is consistent with \cite{Chien2024MNRAS.532..719C}, who uses the JWST CEERS survey that has lower redshifts and less MIRI pointings than SMILES to produce similar results with a smaller sample size of 42 AGNs compared to our 278. And \cite{Wang2020MNRAS.499.4068W} where they used 9 MIR AKARI bands to build an extinction-free census of 126 AGNs in the North Ecliptic Pole-Wide field (NEP). This may suggest that distant galaxies in earlier stages of the Universe were more likely to host AGNs that contribute more to the galaxy's overall IR emissions. In addition, our samples extend to fainter infrared luminosities than those reported by \cite{Wang2020MNRAS.499.4068W}. This extension is likely driven by the improved sensitivity of JWST compared to AKARI. The enhanced sensitivity of JWST enables the detection of significantly fainter galaxies, allowing AGN activity to be identified in regimes that were previously inaccessible. As a result, our analysis can probe AGN populations at lower luminosities and earlier evolutionary stages, including reionization \citep{2015A&A...578A..83G, 2018A&A...613A..44G}.

We also investigate how AGN number fraction changes with redshift. Our results show that AGN number fraction increases with redshift, which again is consistent with previous studies \citep{Wang2020MNRAS.499.4068W,Chien2024MNRAS.532..719C}. \cite{Kirkpatrick2023ApJ...959L...7K} also reports increasing number fraction with redshift, using JWST CEERS and identified AGN through color-color selection process. This might hint at higher, more active populations of AGN residing at higher redshift, earlier in the Universe. Similarly, \cite{2023PASJ...75.1246H} worked with a large sample of 2688 radio (VLA), mid-IR (WISE), and X-ray (XMM-Newton) AGNs, and found that all across the board AGN number fraction increases with redshift. Even smaller sample sizes like the one presented in \cite{2020ApJ...899...35T} using a crossmatched catalog of 583 AKARI and HSC sources found increased AGN activity with increasing redshift (1 $<$ z $<$ 4). These results indicate that AGN are more common in dusty, high redshift populations. Moreover, it is also important to point out \cite{2025A&A...697A.175S} conducted a study similar to ours using JWST JADES without the help of JWST SMILES and reported that number fraction had no dependence on redshift, leading us to believe that the 8 continuous MIRI filters and the increased MIRI pointings in SMILES may play a role in determining the populations of AGNs in different luminous environments

The increase in AGN IR contribution and number fraction with redshift is broadly consistent with the evolution of the cosmic star formation rate density, which peaks at z $\sim$ 2–3 before declining toward the present day \citep{Madau2014ARA&A..52..415M, 2006ApJ...651..142H}. This parallel behavior supports the picture of co-evolving supermassive black holes and their host galaxies, where both star formation and AGN activity are driven by a shared supply of cold gas. Observations show that high-redshift galaxies contain significantly larger gas reservoirs than their local counterparts \citep{2020ARA&A..58..157T}, which can fuel both star formation and black hole accretion. As gas is transported toward the central regions, it can simultaneously sustain star formation and trigger AGN activity, naturally producing the increasing AGN contribution observed in this work.

However, we do not find strong statistical evidence for a systematic dependence of either AGN contribution or AGN number fraction on total IR luminosity, as both trends remain broadly consistent with being flat within the uncertainties. These results appear inconsistent with other works like \cite{10.1093/pasj/psz012} who despite also using AKARI like \cite{Wang2020MNRAS.499.4068W} uses 18 NIR and MIR bands (9 more) and suggests that AGN contribution actually increases with increased luminosity. \cite{10.1093/mnras/stu130} reports finding more AGNs in more IR luminous systems using Herschel-PACS observations in the same GOODS-S file we use in our study. Additionally, \cite{2005MNRAS.360..322G} matched data from IRAS and SDSS resulting in 4248 IRGs, and found that AGN number fraction increases with total IR luminosity. He also shows how [OIII] luminosity and IR luminosity present a good correlation and may be used to find more AGNs. \cite{2010ApJ...721...98K} used a sample of 70$\mu$m selected galaxies from Spitzer and reported that sources identified as AGN constitute over 70\% of sources at high luminosity (log($L/L_{\odot}$) $>$ 12). These differences may arise from a combination of sample selection, AGN identification techniques, wavelength coverage, and the improved sensitivity of JWST, which enables the detection of fainter and more heavily obscured AGN populations than were accessible to previous infrared surveys.

While the flat number fraction and AGN contribution trend initially appears at odds with studies that report increasing AGN incidence with infrared luminosity \citep{2024ApJ...966..229L}, it can be physically explained by the non-linear relation between mid-IR and total IR dominance outlined by \cite{2015ApJ...814....9K}. AGN-heated dust peaks strongly in the rest-frame near-to-mid-IR ($2\text{--}30\,\mu\text{m}$) window where our deep MIRI observations are highly sensitive. Conversely, $L_{\text{TIR}}$ integrates the entire $8\text{--}1000\,\mu\text{m}$ spectrum, which is heavily dominated at longer, far-IR wavelengths ($\lambda \gtrsim 100\,\mu\text{m}$) by cold dust heated by intense star formation. Consequently, an AGN can contribute significantly to the MIRI bands while representing only a modest fraction of the total integrated $L_{\text{TIR}}$ energy budget if hosted within a powerful starburst system. This may explain why we do not observe a strong relationship between AGN activity and total IR luminosity.

Moreover, our ability to constrain any luminosity dependence at the highest luminosities is limited by the small number of sources with log($L/L_{\odot}$) $>$ 12 luminosities in our sample. In this regime, previous studies focusing on more luminous systems provide valuable complementary constraints.

Although we do not observe a strong luminosity dependence, several physical and observational effects may contribute to the relatively flat behavior. Star formation and AGN activity are not expected to evolve simultaneously, and the stochastic nature of black hole accretion can produce galaxies with similar total infrared luminosities but differing AGN contributions \citep{2009ApJS..182..628V, 2015MNRAS.453..591S}. In addition, because redshift and luminosity are inherently correlated in flux-limited surveys, higher-redshift bins are preferentially populated by intrinsically luminous galaxies, while lower-redshift bins span a wider luminosity range. These selection effects may partially obscure any intrinsic dependence of AGN activity on total infrared luminosity.

Our results provide new insights into the populations of AGNs, particularly in high redshift galaxies where obscured AGNs are difficult to detect. While our findings align with some previous works, the superior continuous mid-IR coverage of JWST enables a more detailed understanding of the relationship between AGN activity and host galaxy properties. Despite these advantages, our study is limited by the use of photometric redshifts, which may introduce uncertainties, especially when conducting SED fitting. Furthermore, although our sample includes 2,735 galaxies, it is limited by the current depth of infrared observations, meaning that some faint AGNs may still be missed.

Our findings highlight the crucial role of mid-IR observations in the study of AGNs, and suggest that JWST is advancing our understanding of AGN and galaxy evolution. However, future work will benefit from deeper IR surveys and improved spectroscopic redshifts. These will enable more precise measurements of AGN properties and their evolution, especially at higher redshifts. As JWST continues to observe the distant universe, future studies will be key in refining our models of AGN activity and its impact on galaxy evolution.

\section{Conclusion} \label{sec:Conclusion}

In this investigation, we focused on examining the evolution of AGN infrared contribution and number fraction as a function of redshift and total infrared luminosity, utilizing data from a  merged catalog of the SMILES and JADES datasets from JWST. Our sample includes 2735 galaxies from the CIGALE SED fitting results performed by \cite{2026PASP..138a4102L} to study the the physical properties of AGN. The continuous 8-band MIRI coverage of SMILES enables more robust SED fitting and a cleaner separation between AGN and star-forming emission than studies relying on more limited mid-infrared coverage. Compared to earlier JWST surveys such as CEERS, the depth and wavelength continuity of SMILES allow us to identify a substantially larger and more complete AGN sample, strengthening the statistical significance of these results. 

We find that both AGN contribution and AGN number fraction increase with redshift (Figure \ref{subfig:AGN_contA}  and Figure \ref{subfig:AGN_numfracA}), consistent with previous studies. In contrast, neither AGN contribution nor AGN number fraction shows strong statistical evidence for a systematic dependence on total infrared luminosity, with both trends remaining broadly consistent with being flat within the uncertainties (Figure \ref{subfig:AGN_contB} and Figure \ref{subfig:AGN_numfracB}).

These results suggest that obscured AGN activity becomes more prevalent at higher redshift, while showing little evidence for a strong dependence on total infrared luminosity. This indicates that, within our sample, cosmic epoch may play a more important role than total infrared luminosity in shaping the prevalence and relative importance of obscured AGN.

Overall, these findings highlight the importance of deep, continuous mid-infrared observations with JWST for revealing dust-obscured AGN populations, and show a glimpse of the advances we have made in our understanding of galaxy and AGN evolution through cosmic time.

\begin{acknowledgments}
The authors are grateful to the anonymous referee for the valuable comments, which significantly improved the paper.

This research is based on observations made with the NASA/ESA/CSA James Webb Space Telescope. The data were obtained from the Mikulski Archive for Space Telescopes (MAST) at the Space Telescope Science Institute, which is operated by AURA under NASA contract NAS 5-03127. We thank the JWST SMILES and JADES teams for making their data publicly available, which made this work possible.

T.G. acknowledges the support of the National Science and Technology Council of Taiwan through grants 114-2112-M-007 -026, 113-2112-M-007 -006, and 113 -2927-I-007 -501.
T.H acknowledges the support of the National Science and Technology Council of Taiwan through grants 110-2112-M-005 -013 -MY3 and 110-2112-M-007 -034. 

A.R.B. acknowledges support from the Institute of Astronomy at National Tsing Hua University. We also thank our collaborators and colleagues for valuable discussions and feedback that improved this work.
\end{acknowledgments}

\facilities{HST, JWST}

\software{astropy \citep{astropy:2013, astropy:2018, astropy:2022},  
          CIGALE \citep{Boquien2019A&A...622A.103B}, 
          EAZY \citep{Brammer2008ApJ...686.1503B}
          }

\bibliography{PASPsample701}{}
\bibliographystyle{aasjournalv7}

\end{document}